\documentclass[10pt, conference]{IEEEtran}
\IEEEoverridecommandlockouts

\usepackage{cite}

\ifCLASSINFOpdf
  \usepackage[pdftex]{graphicx}
\else
\fi

\usepackage{graphicx}
\usepackage{amssymb}
\usepackage{xcolor}
\usepackage{soul}
\usepackage{amsmath}
\usepackage{epsfig}
\usepackage{comment}
\usepackage{subfigure}
\usepackage{subcaption}
\usepackage{multirow}
\usepackage{makecell}
\usepackage{array}

\usepackage{url}

\usepackage[lofdepth,lotdepth]{subfig}
\usepackage{float}
\usepackage[top=0.75in, bottom=1.05in, left=0.65in, right=0.65in]{geometry}

\definecolor{myyellow}{rgb}{1, 1, 0.77}

\begin{document}

\title{Experimental Performance of a 5G N78 Reconfigurable Intelligent Surface: From Controlled Measurements to Commercial Network Deployment}

\author{
\IEEEauthorblockN{
Sefa Kayraklık\textsuperscript{$\ast$},
Samed Keşir\textsuperscript{$\circ$},
Batuhan Kaplan\textsuperscript{$\circ$},\\
Ahmet Muaz Aktaş\textsuperscript{$\ast$},
Emre Arslan\textsuperscript{$\circ$},
Ahmet Faruk Coskun\textsuperscript{$\circ$}
}

\IEEEauthorblockA{\textsuperscript{$\ast$}Communications and Signal Processing Research (HISAR) Lab., TUBITAK BILGEM, Kocaeli, Turkiye}

\IEEEauthorblockA{\textsuperscript{$\circ$}6G and Artificial Intelligence Laboratory, Turkcell Iletisim Hizmetleri Inc., Istanbul, Turkiye}
    
\IEEEauthorblockA{Emails: \textsuperscript{$\ast$}\{sefa.kayraklik,muaz.aktas\}@tubitak.gov.tr\\
\textsuperscript{$\circ$}\{samed.kesir,batuhan.kaplan,emre.arslan,coskun.ahmet\}@turkcell.com.tr
}}

\maketitle

\begin{abstract}
This paper presents a real-world experimental analysis of a modular reconfigurable intelligent surface (RIS) prototype designed to operate in the 5G N78 band. Unlike most RIS studies in the literature that focus on simulations or controlled setups, the proposed system is validated through three phases consisting of indoor measurements, outdoor long-range tests, and deployment in a live commercial 5G standalone network. The RIS is exploited to enhance coverage in a non-line-of-sight (NLoS) zone, identified through baseline drive tests. Results show promising gains in RSRP and SINR, while also restoring 5G service at user locations where access was previously not available. The results highlight the practical potential of RIS for coverage enhancement in operational 5G networks.  
\end{abstract}


\section{Introduction}
With the deployment of 5G in many countries, the wireless communications industry is rapidly moving from 5G towards the initial planning of 6G. Due to its specific characteristics in providing a good balance with fast data speeds while still having extensive coverage, the N78 spectrum (operating between the 3.3 and 3.8 GHz range) is attractive to modern wireless networks \cite{111_3gpp38104}. However, N78 signals have a major physical issue that needs to be addressed. Although they penetrate buildings better than higher frequency signals, they are still easily blocked by large concrete walls and buildings, which leads to coverage gaps and dead zones where the signal cannot reach the user or becomes very weak. Conventional solutions to fix these coverage gaps have usually been based on deploying additional base stations (BSs), leading to major expenses for operators due to high energy costs, hardware installation requirements, and site rentals in crowded cities.

Reconfigurable intelligent surfaces (RISs) are a potential solution to overcome these environmental issues \cite{222_Basar2019RIS,333_Wu2021RISTutorial}. 
Due to their semi-passive properties, RISs can extend the coverage without massive power consumption or complex wiring and set-ups that is required from traditional active cellular towers. The benefits of RIS are well established in the literature; however, most studies depend on computer simulations and assumptions. Only a limited number of studies include real-world field trials, leading to a gap in practical analysis and realistic performance evaluation of RIS technology \cite{444_Dai2020RISPrototype,555_Araghi2022Sub6RIS}. 
Besides, while the potential of RISs has been proven and tested at higher frequencies, their performance in the 5G mid-band is much harder to measure due to the way signals scatter in natural settings \cite{333_Wu2021RISTutorial,444_Dai2020RISPrototype,555_Araghi2022Sub6RIS}. This lack of practical and experimental studies from live network infrastructures is the biggest barrier preventing the deployment of RIS in future networks and consideration in future global 6G standards \cite{EA_Practical_2, 555_Araghi2022Sub6RIS}.

The aim of this study is to fill this gap, providing an extensive three-phase evaluation of the RIS prototype that is optimized for the N78 frequency band \cite{yerliRIS}. Our primary contributions focus on the systematic testing methodology that moves from controlled indoor lab experiments to outdoor long-distance trials. Finally, we provide experimental validation within a live commercial 5G network, exploiting Turkcell’s infrastructure to identify real-world coverage gaps and show actual performance improvements in areas where the signal was previously blocked. The specific key contributions of this study are summarized as follows. First, we establish a performance baseline through controlled indoor and outdoor experiments to isolate and measure the exact gain provided by the RIS prototype. Secondly, we conduct a baseline drive-test campaign using a commercial receiver to map coverage holes caused by building blockages in a live testbed. Finally, we measure and quantify the RIS-induced improvements in signal strength-based key performance indicators (KPIs) such as the reference signal received power (RSRP), and signal to interference plus noise ratio (SINR), going beyond simulation and controlled proof-of-concept testing by validating RIS performance within a real-life commercial 5G network. Hence, we demonstrate that RIS technology can restore high-quality service to locations in a realistic scenario where maintaining a connection was previously impossible.

The remainder of the paper is organized as follows. Section II introduces the overview of the RIS-assisted networks, the specifications of the RIS prototype and the utilized test infrastructure as well as the propagation characteristics of the 5G N78 band. Section III provides the detailed descriptions and methodology of each experimentation phase. The corresponding experimental results obtained are exhibited in Section IV. Finally, Section V concludes the paper by summarizing key insights and outlining directions for future research.

\section{Background and System Overview}
\vspace{0.08in}
This section provides the necessary background and system context. It first introduces the operating principles of RIS and the N78-compatible prototype used in this study, then discusses the propagation characteristics of the 5G N78 band, and finally outlines the commercial 5G network architecture and the KPIs considered.

\subsection{RIS Fundamentals and Prototype Specifications}
Using programmable reflective elements, RISs can steer reflected electromagnetic waves toward desired directions. 
The signal reflected by $N$-element passive RIS can be given as $y = \left(\sum_{i=1}^{N} h_i e^{j\phi_i} g_i \right) x + n$,
$x$ is the transmitted signal, $\phi_i$ is RIS phase shift configurations, $h_i = a_ie^{-j\theta_i}$ and $g_i = b_ie^{-j\psi_i}$ are channel coefficients of transmitter-RIS and RIS-receiver, respectively. $n$ is additive white Gaussian noise. 
To maximize the received instantaneous signal-to-noise ratio (SNR), the phase of each RIS element should be set as $\phi_i = \theta_i + \psi_i$ for all $i = 1 \dots N$, assuming continuous phase adjustment. 

\begin{figure}[t]
    \centering    
    \includegraphics[width=0.75\columnwidth]{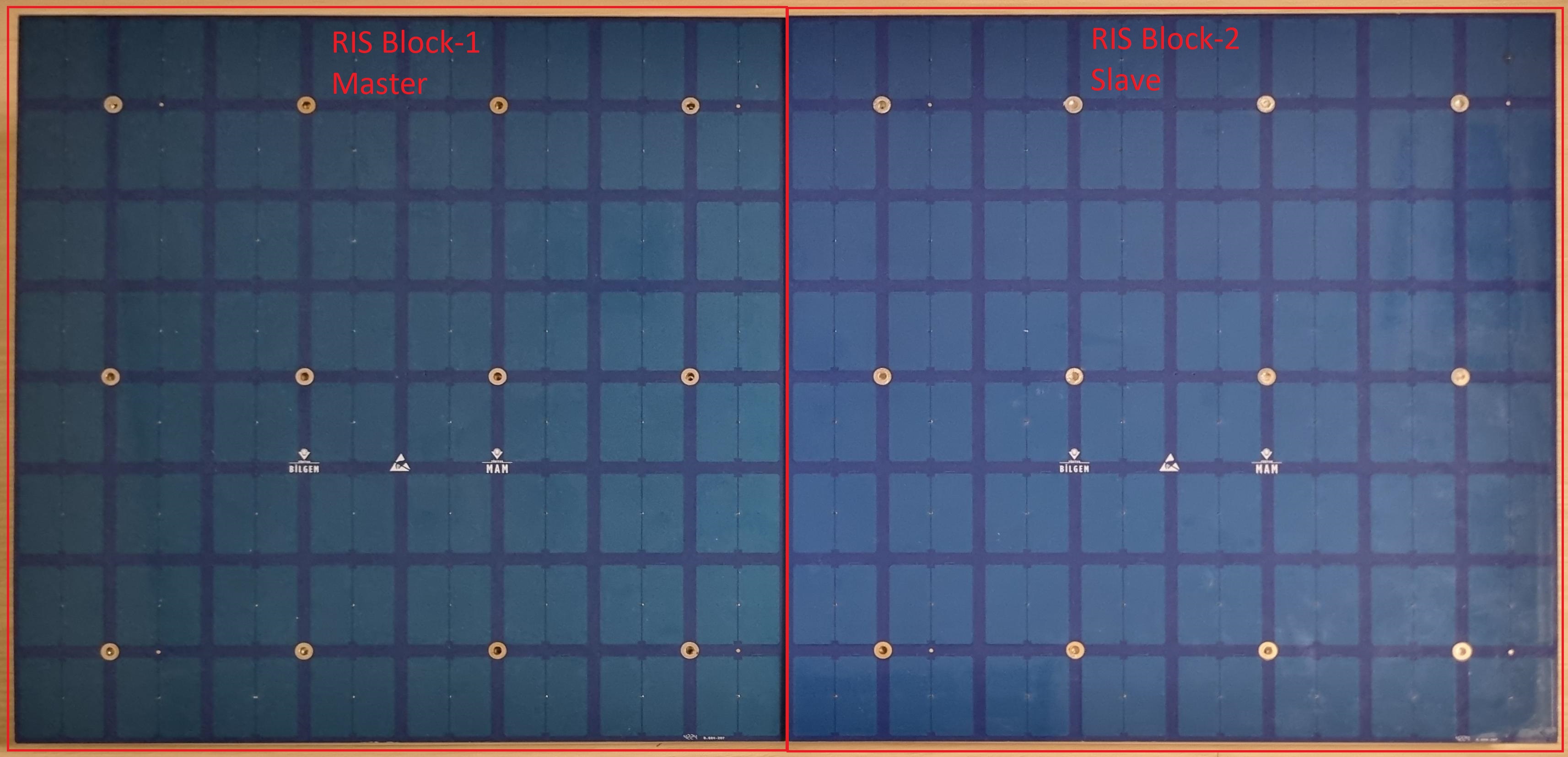}
    \caption{N78 RIS prototype structure}\vspace{-10pt}
    \label{fig:RIS}
\end{figure}

The RIS prototype \cite{yerliRIS} developed by TUBITAK, which is shown in Fig. \ref{fig:RIS}, utilizes a meta-surface structure that is composed of four metallic layers and three dielectric substrate layers. The first dielectric layer uses a $3$ mm-thick F4BM substrate, while the remaining two are $0.5$ mm-thick FR4 layers. On the top metallic layer, two circular-aperture patches are connected by a PIN diode, with microstrip line extensions for DC biasing. The RIS contains reflecting elements operating in the N78 frequency band, whose phase response is controlled by a PIN diode to achieve a $180^\circ$ phase difference between the diode's \emph{on} and \emph{off} states. The hardware architecture is highly scalable, featuring a master controller and up to sixteen interchangeable slave blocks. Each block shares an identical hardware structure with an $8\times 8$ array of reflecting elements ($64$ elements total) and a dedicated controller board. A microcontroller supporting Wi-Fi, I2C, UART, and USB interfaces drives self-configuration and inter-block communication, routing logic from the rear-mounted control ICs to the front-facing PIN diodes, enabling wireless and wired configuration of the RIS. Regarding physical dimensions and deployment configuration, each unit reflecting element has a periodicity of $41$ mm. The modular deployment configuration allows multiple $8\times8$ blocks to be seamlessly tiled together to form large-scale reflecting surfaces (e.g., combined $1\times2$, $2\times2$, and $3\times3$ block layouts). 

\subsection{Channel Characteristics of 5G N78 (C-Band)}
This band, operating in the $3.3$--$3.8$ GHz range, represents a key mid-band spectrum providing a balance between coverage and capacity. Compared to mmWave, N78 signals experience lower path loss and improved penetration, enabling wider connectivity and more robust performance in practical cases \cite{bjornson2019mMIMO, etsiRIS2025}. The propagation environment at C-band is typically rich in scattering, with multiple reflections, and diffractions contributing to more stable signal reception, and the path loss exponent in line-of-sight (LoS) conditions varies between $1.6$ and $2.4$ depending on the scenario, reflecting near free-space propagation. In contrast, most RIS research has focused on mmWave, where channels are sparse, LoS conditions dominate, and severe path loss makes RIS particularly effective for enabling alternative NLoS paths and \cite{bjornson2019mMIMO, wu2019irs}. In the N78 band, the path loss exponent remains relatively moderate even in NLoS conditions compared to mmWave, implying that the inherently richer scattering and limited attenuation might reduce the relative impact of RIS, potentially limiting its performance gains. Meaningful improvements are generally observed only under strong blockage or geometrically constrained environments, making experimental demonstrations in operational networks more challenging ~\cite{sang2023sub6, 555_Araghi2022Sub6RIS}.

Despite these challenges, evaluating RIS in the N78 band is highly relevant due to its wide adoption in commercial 5G networks. Even in this mid-band, RIS can provide tangible benefits, such as enhancing coverage in shadowed regions, improving signal quality at cell edges, and increasing link reliability under partial blockage~\cite{kilinc2021channel, wu2019irs}. These observations highlight the practical potential of RIS as a complementary technology to improve service quality and coverage in real-world 5G N78 deployments, particularly in dense urban areas where additional 5G site deployment would be costly or challenging due to high requirements for fiber or high-speed transport, increased energy demand, and site rental constraints.

\subsection{Commercial 5G Network Architecture}
The 5G network architecture consists of Ericsson's Active Antenna Unit (AAU) and Base Band Unit (BBU) products, core network solutions provided by i2i Systems, and a General Mobile OD513 customer premises equipment (CPE). AAU serves as the key transmission element and operates in an N78 band with a channel bandwidth of $100$ MHz. Also, AAU operates in a single-beam mode with no specified SSB beam dedicated the RIS. 
Connected to the AAU, the BBU acts as the central processing unit of the network, where signal processing and resource management are executed. The CPE acts as a UE that establishes a connection to the BS. General Mobile CPE is utilized on the receiver side, which is equipped with the Qualcomm Snapdragon X55 5G chipset. During the measurement, KPI records are monitored for the RIS-on and RIS-off cases. Among the various KPIs, Received Signal Strength Indicator (RSSI), RSRP, Reference Signal Received Quality (RSRQ), and SINR are collected to demonstrate the efficiency of the 5G NR signal reflected via the RIS. 

\section{Measurement Methodology}
This section describes the experimental methodology adopted to perform the three-phase performance evaluation of the RIS prototype. The first phase involves controlled indoor tests, the second phase covers outdoor links at increased distances, and the third phase evaluates RIS-assisted performance over a commercial 5G network at selected user locations.

\subsection{Phase-1: Indoor Controlled Measurements}
\begin{figure}[t]
  \centering
  \subfigure[Measurement environment]{\includegraphics[width=0.23\textwidth]{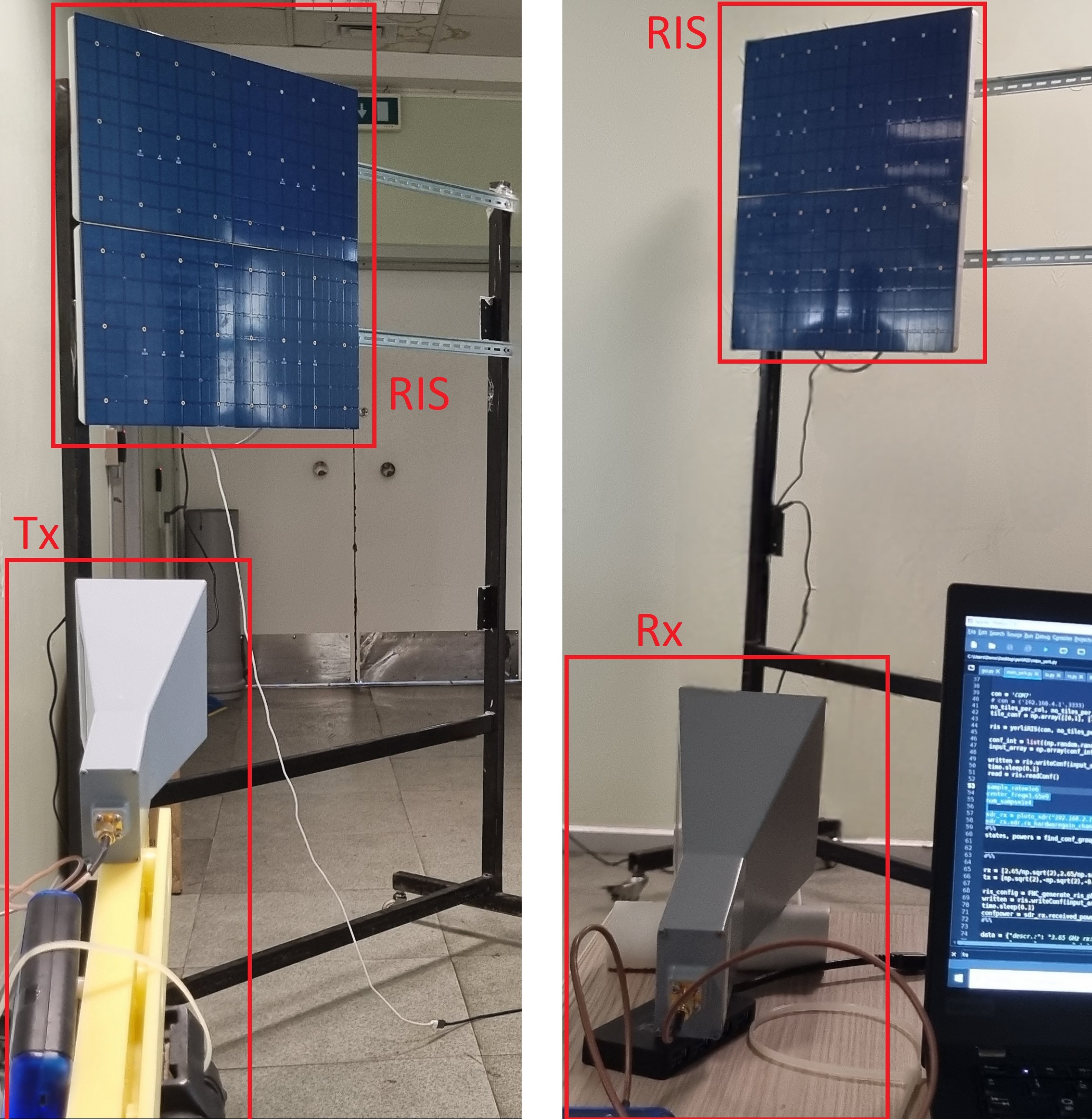}} \hfill
  \subfigure[Experiment scenarios]{\includegraphics[width=0.24\textwidth]{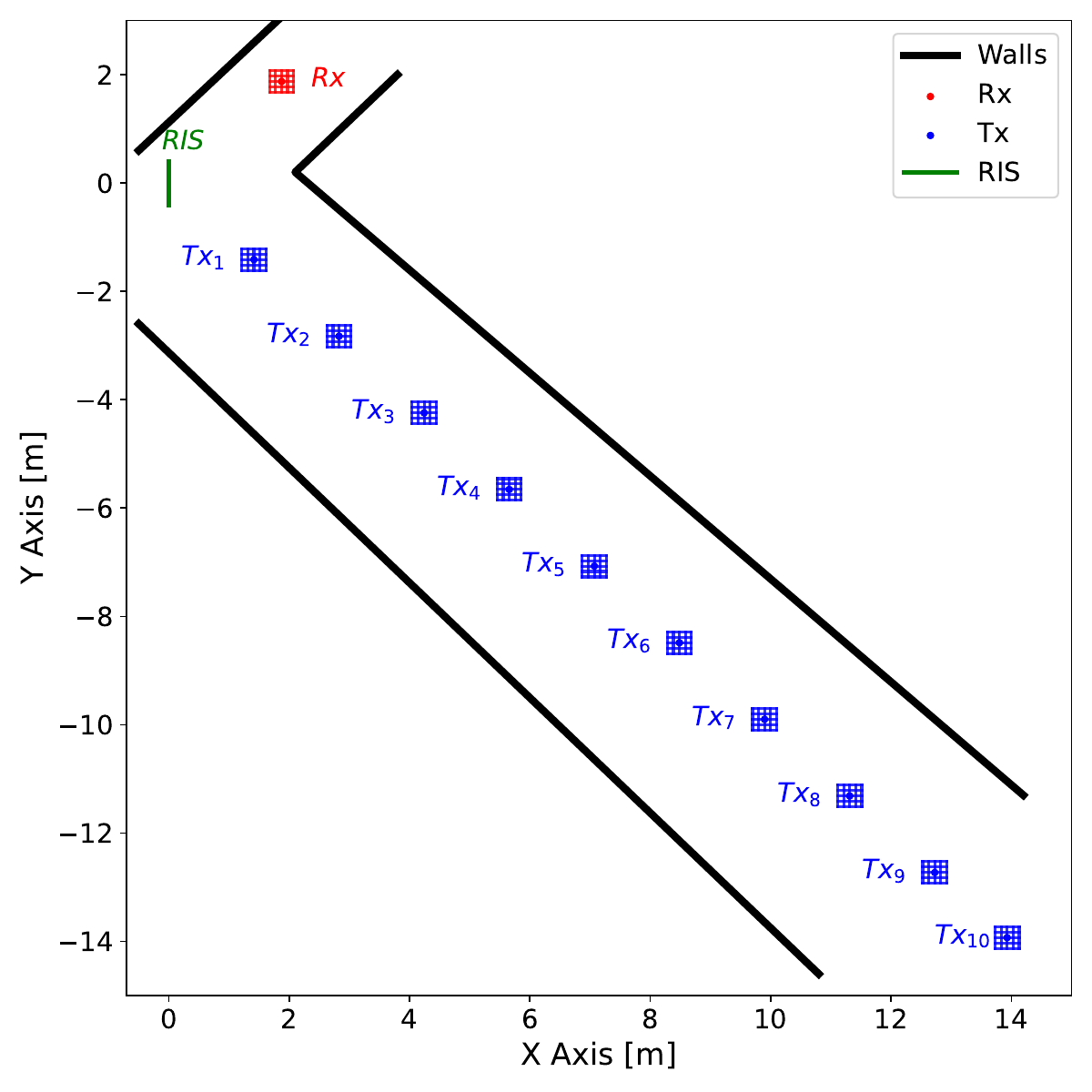}}
  \caption{Indoor RIS-assisted communication}\vspace{-10pt}
  \label{fig:indoor_setup}
\end{figure}

This phase aims to establish a baseline understanding of the achievable enhancement by isolating the system from external environmental uncertainties. To this end, a measurement setup based on software-defined radios (SDRs) and horn antennas is employed to emulate the transceivers. The controlled nature of this setup enables precise characterization of the RIS-induced gain and serves as a reference for subsequent outdoor and real-world measurements. The major components of the measurement setup illustrated in Fig. \ref{fig:indoor_setup}-(a) might be listed as:
\begin{itemize}
    \item Adalm Pluto SDR: Supports the frequency range of $[325, 3800]$ MHz with $20$ MHz instant bandwidth.
    \item Horn antennas: TUBITAK horn antenna features a $40^\circ$ half-power beamwidth in sub-6~GHz with $13$ dBi gain.
\end{itemize}

Indoor experiments were performed using a single-input single-output configuration with Adalm Pluto SDRs serving as both transmitter and receiver. There are $256$ reflecting elements, composed of four RIS blocks, positioned $2.83$ m from the receiver. Meanwhile, the transmitter is deployed at ten equally spaced locations relative to the RIS, with the corresponding distances shown in Fig. \ref{fig:indoor_setup}-(b). Iterative \cite{iterative}, and location-based \cite{BFTypeRISConfig} algorithms are employed to optimize the RIS configuration. For both approaches, the performance metric is defined as the received signal power measured at the SDR. The iterative algorithm sequentially tests each element’s possible states and selects the one that maximizes received signal power, repeating this process element by element. On the other hand, the location-based algorithm determines the phase shift of each RIS reflecting element by calculating the path loss associated with its corresponding channel coefficients. In the location-based approach, since the exact coordinates of the transmitter and receiver are unknown, a grid search algorithm is utilized to refine the positional estimates. The algorithm exhaustively tests all potential points around the predicted locations, recording the RIS configuration corresponding to the transmitter-receiver position pairs that provide the highest received power levels. 

\begin{figure}[t]
  \centering
  \subfigure[Measurement environment]{\includegraphics[width=0.49\textwidth]{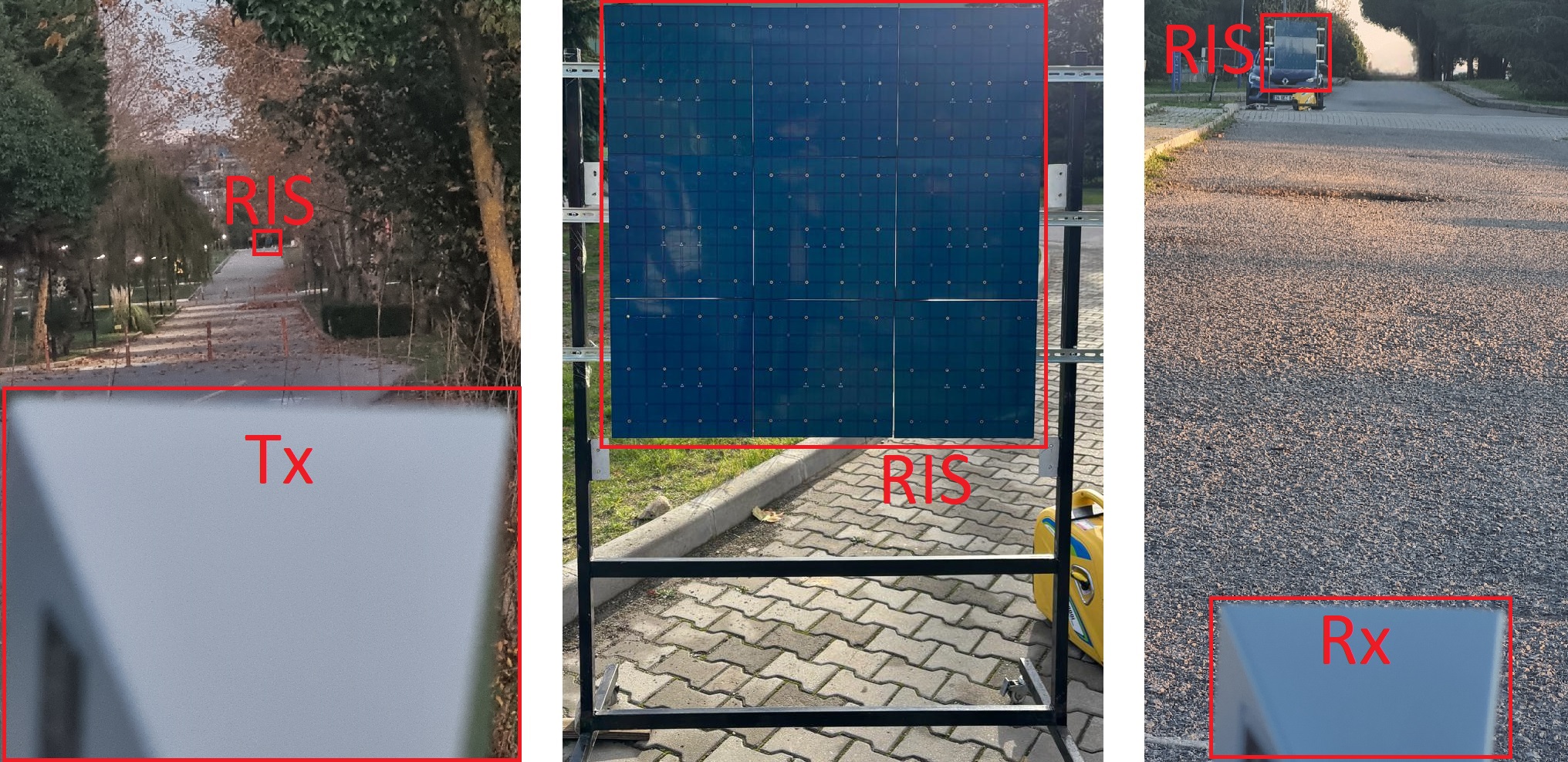}} \hfill
  \subfigure[Experiment scenarios]{\includegraphics[width=0.49\textwidth]{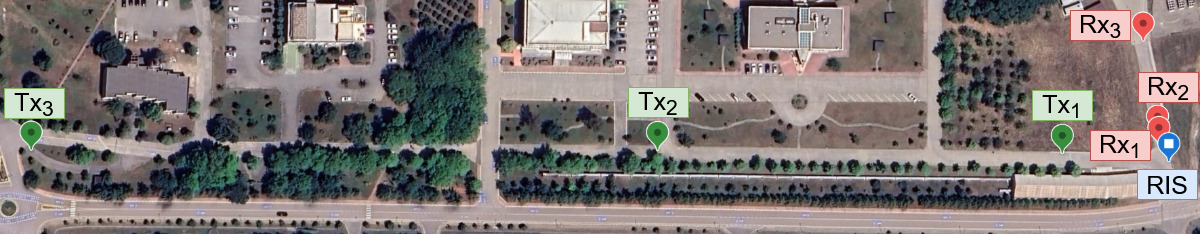}}
  \caption{Outdoor RIS-assisted communication}\vspace{-10pt}
  \label{fig:outdoor_setup}
\end{figure}

\subsection{Phase-2: Outdoor Controlled Measurements} 
The primary objective is to investigate the performance of the RIS under increased link distances and more realistic propagation conditions, including multipath effects and environmental variability. Similar to the first phase, measurements are conducted using SDR-based transceivers and directive antennas to ensure consistency in the experimental setup. 
Fig. \ref{fig:outdoor_setup} illustrates the measurement environment and experimental scenarios used to evaluate the performance of the RIS-assisted wireless communication outdoors. Similarly, the measurement campaign consists of SDRs with horn antennas to transmit and receive the over-the-air signal and a larger RIS configuration of $576$ reflecting elements, composed of nine blocks to increase the reflection aperture, whose environment is shown in Fig. \ref{fig:outdoor_setup}-(a). The experimental geometries, depicted in Fig. \ref{fig:outdoor_setup}-(b), include three transmitter locations ($50$ m, $225$ m, and $500$ m from the RIS) and three receiver locations ($8$ m, $13$ m, and $55$ m from the RIS) to thoroughly assess outdoor performance. Similar to the previous phase, the iterative and location-based practical RIS optimization methods are utilized to optimize the received signal power by adjusting the RIS phase shifts. Furthermore, for long-distance scenarios where the power contribution of a single reflecting element is indistinguishable, a grouped iterative approach is adopted. By clustering the RIS elements into sets of four, the observable change in received power during each optimization iteration is enhanced.

\subsection{Phase-3: Commercial 5G Network Measurements}
In the final phase of the measurement-based investigation, the RIS prototype will be tested in a commercial-grade 5G wireless network. For this purpose, Turkcell’s 5G standalone (SA) test infrastructure deployed at TUBITAK’s Gebze campus will be utilized. A brief overview of the radio access and core network components is provided in Section II-C. Unlike the previous two measurement phases, where both (i) near-perfect NLoS conditions between the transmitter and receiver nodes, and (ii) the orientation and electromagnetic selectivity factors enabling RIS-assisted communication were ensured in a controlled manner, this phase considers a realistic scenario in which neither of these conditions can be guaranteed. Instead, both factors inherently and uncontrollably vary depending on the deployment environment. As described in Section II-B, due to rich scattering effects, reasonably good communication conditions may still be observed even in the absence of RIS.

This phase consists of two main tasks: (i) identifying coverage gaps under existing non-RIS conditions, and (ii) generating optimized RIS configurations for the identified vulnerable regions and quantifying the RIS-induced performance gains.

\subsubsection{Baseline Coverage Assessment (RIS-Free)}
\begin{figure}[t]
  \centering
  \subfigure[Google Earth view]{\includegraphics[width=0.23\textwidth]{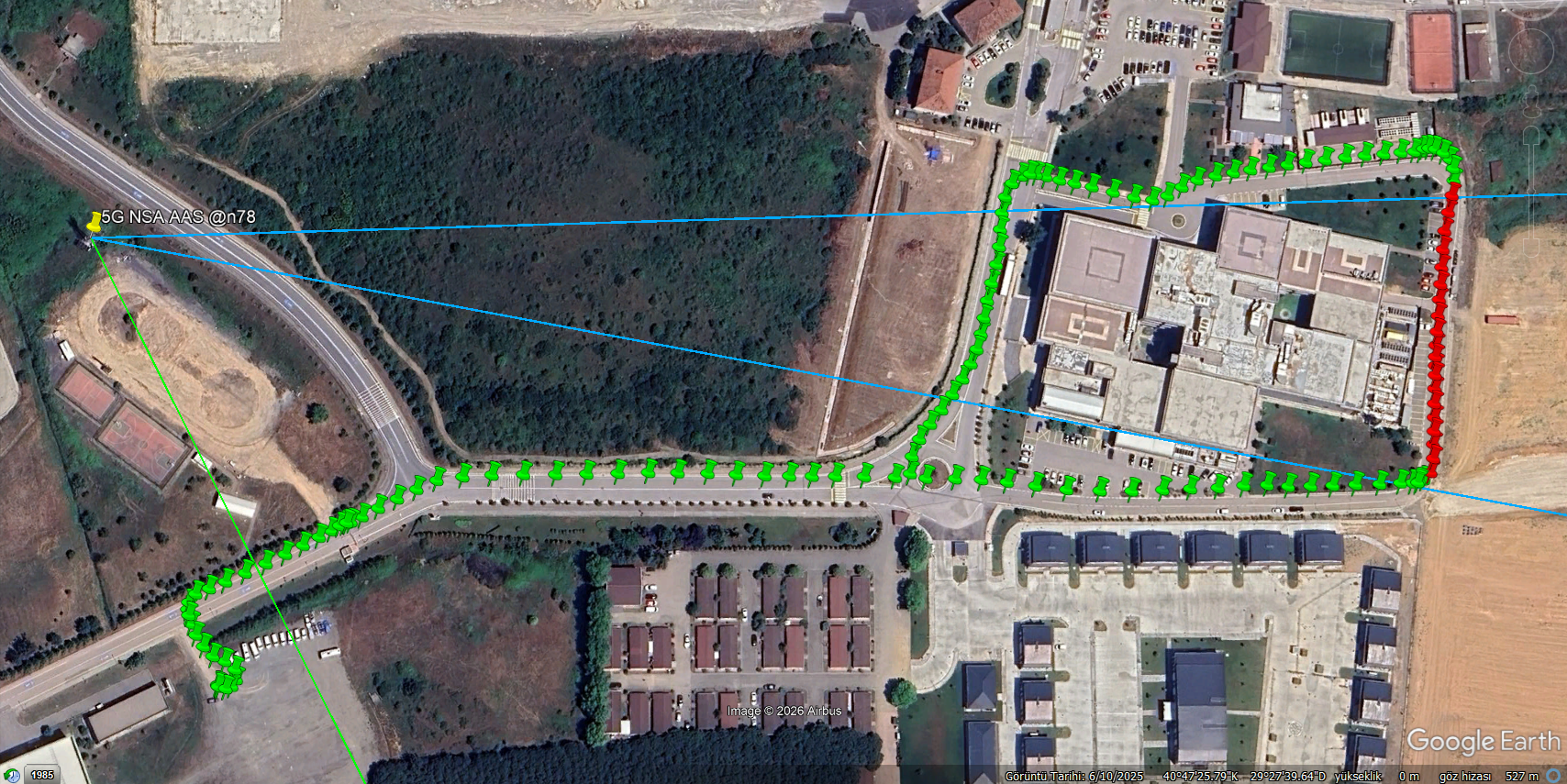}} \hfill
  \subfigure[Simplified representation]{\includegraphics[width=0.25\textwidth]{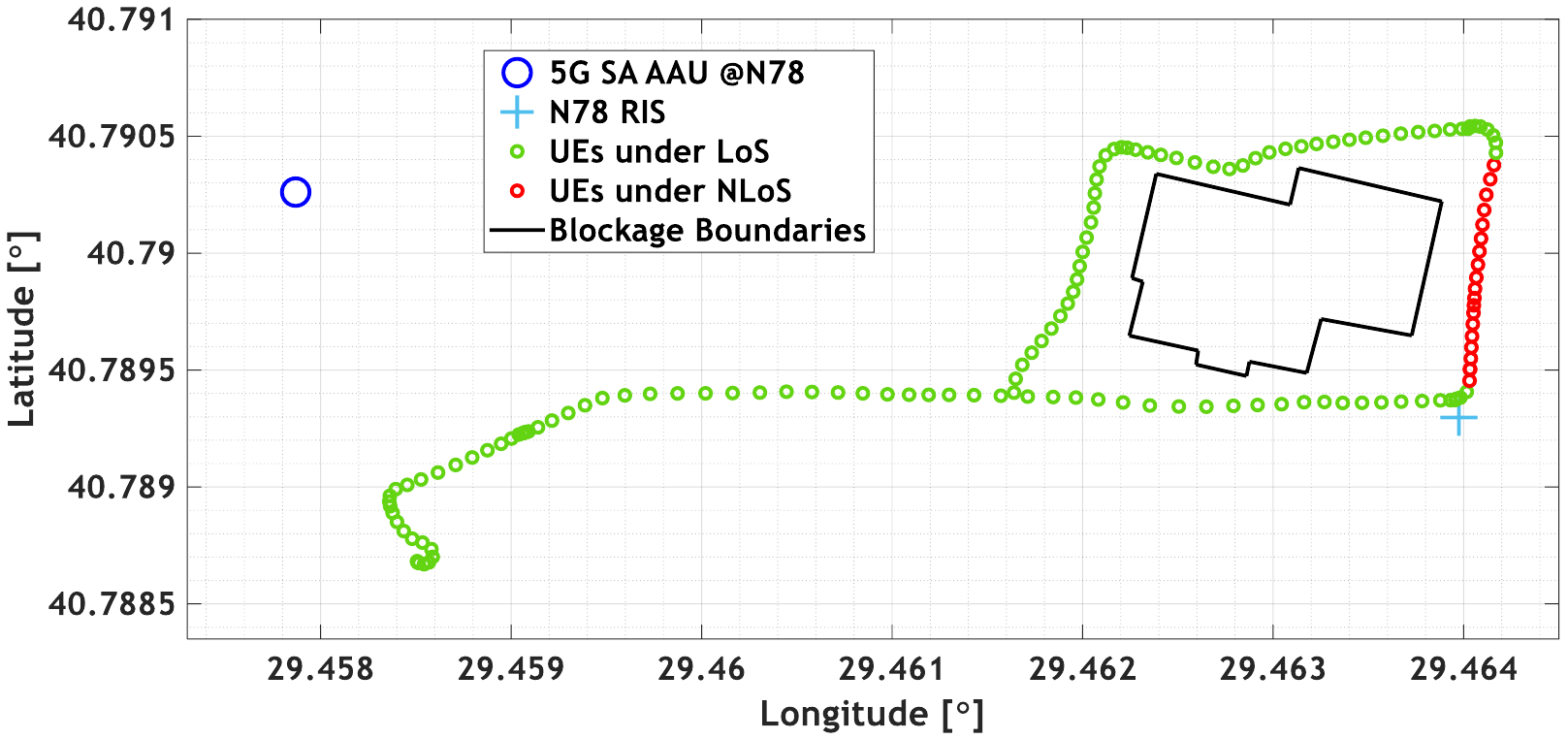}}
  \caption{Top view of the RIS-free site survey showing the network nodes and the UE points with potential reduced signal quality (due to NLoS w.r.t AAU)}\vspace{-10pt}
  \label{fig:N78CellularSetup}
\end{figure}

As the first step, a baseline coverage assessment is performed without RIS deployment to identify UE locations experiencing degraded signal conditions under the serving gNB. By conducting a drive test campaign across the target area without the RIS deployment, the CPE was driven along predefined routes depicted in Fig. \ref{fig:N78CellularSetup} while logging the KPIs, including RSRP, RSRQ, RSSI, and SINR. By monitoring the characteristics of the KPIs experienced by the CPE, particularly in the region located behind the (three-story) building with respect to the gNB LoS direction, coverage weaknesses in this area are identified. In Fig. \ref{fig:N78CellularSetup}-(a), the boresight direction of the AAU at the gNB is illustrated with a green line. The CPE/UE locations considered during the site survey that have a direct optical LoS with the AAU are indicated with green markers, whereas the points located behind the building, where 5G service degradation is expected, are visualized as red markers within the region bounded by the blue lines.

To detect the coverage gaps, the variation of the KPIs collected through drive tests is taken into account together with the well-known free-space path-loss (FSPL) model. A comparison between the measured KPIs and the FSPL estimations would help to identify the coverage gaps caused by blockage rather than natural propagation-based attenuation.

\subsubsection{RIS-Assisted Performance Evaluation Methodology}
\begin{figure}[t]
  \centering
  \subfigure[BS]{\includegraphics[width=0.1100\columnwidth]{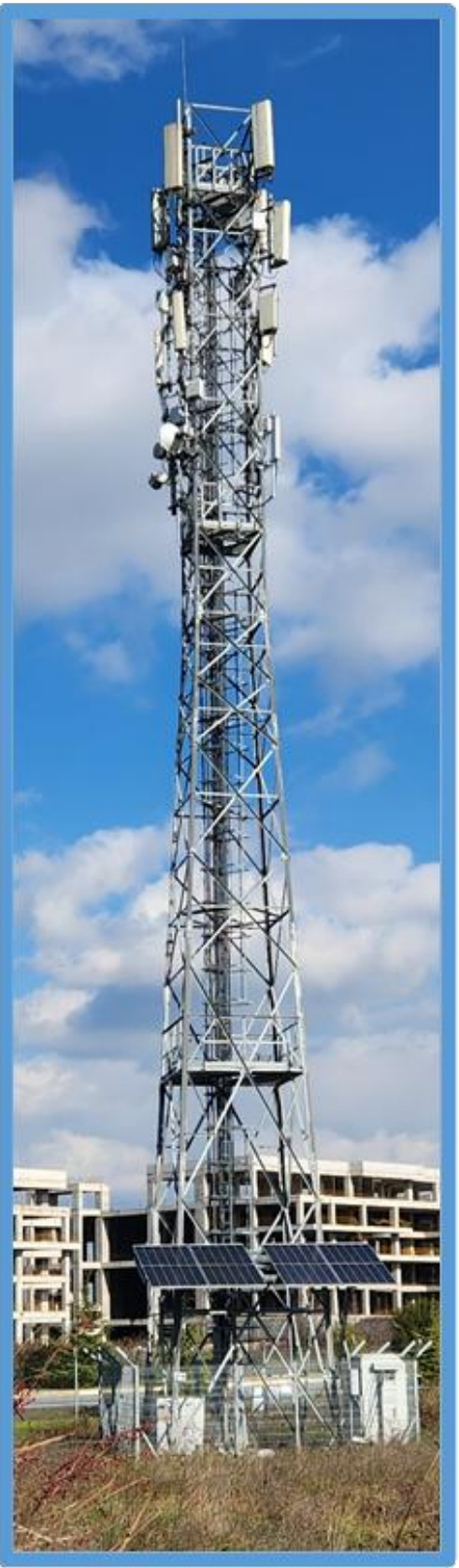}}
  \subfigure[Main scene]{\includegraphics[width=0.6589\columnwidth]{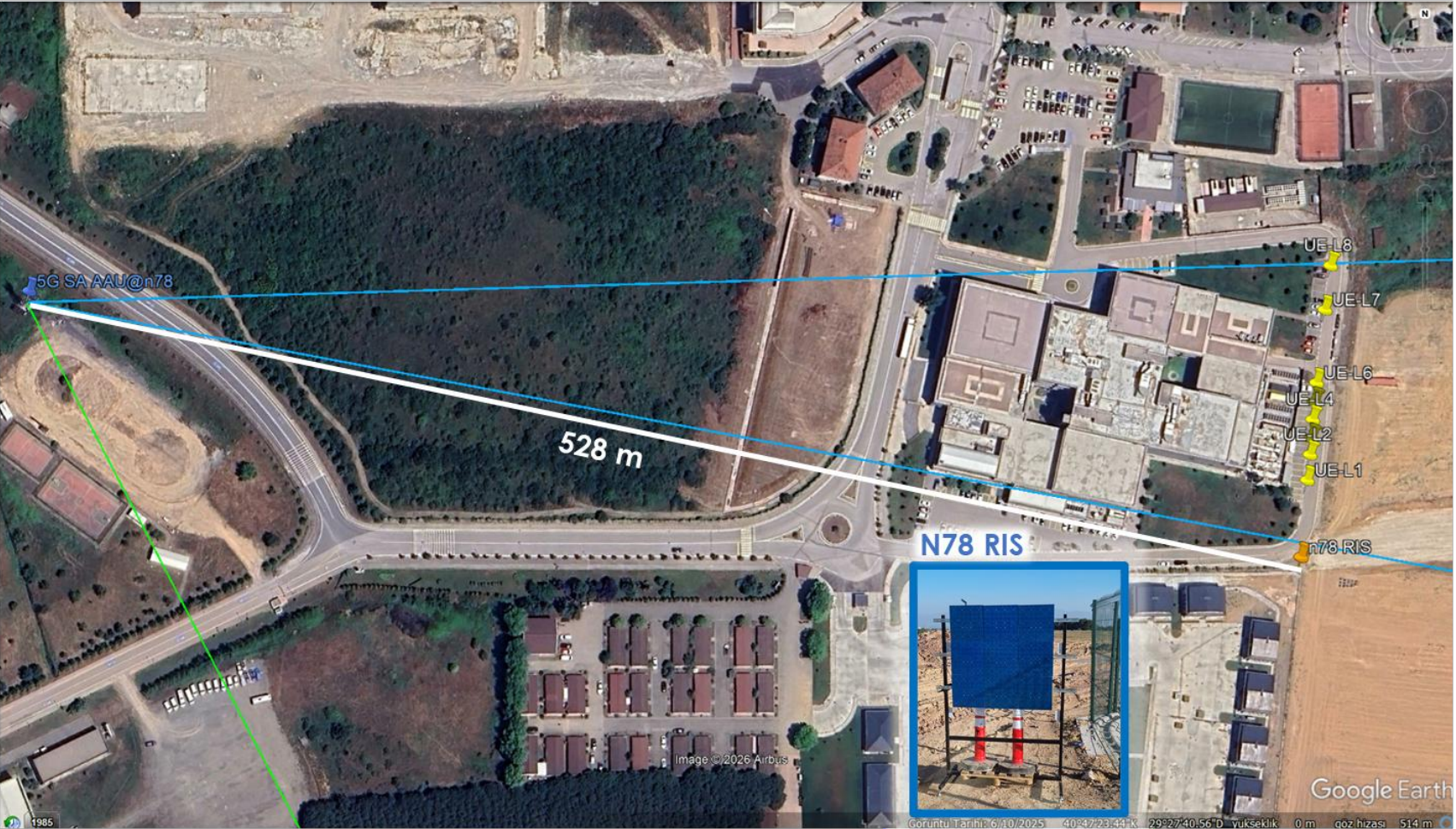}}
  \subfigure[UEs]{\includegraphics[width=0.1827\columnwidth]{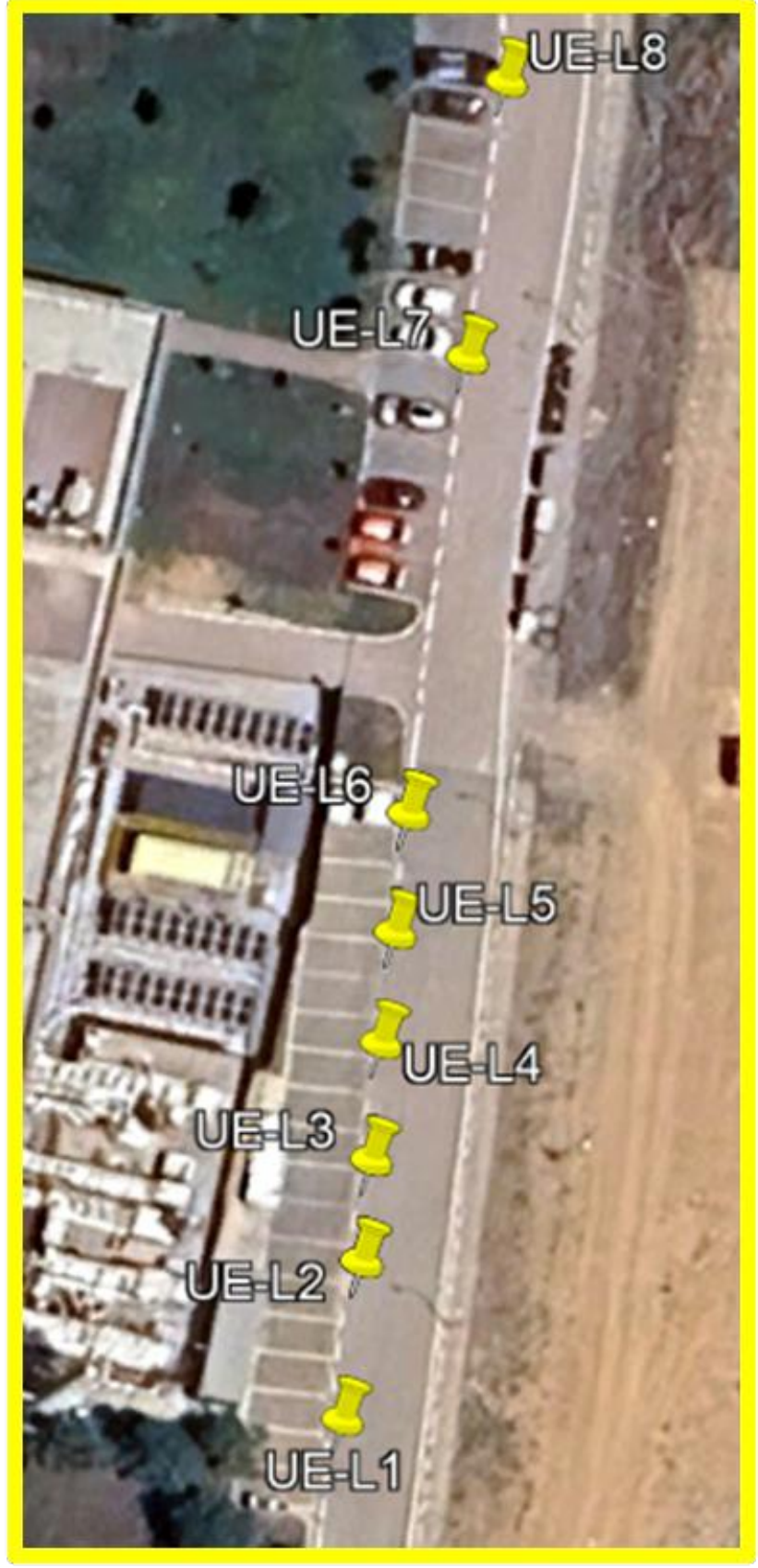}}
  \caption{Top view of the RIS-assisted 5G wireless communications scenario}
  \label{fig:N78RISAssistedSetup}\vspace{-10pt}
\end{figure}

Based on the identified NLoS region behind the major blocking building, the RIS was mounted on a dedicated outdoor stand at a location providing unobstructed LoS to the serving AAU, positioned to redirect the incident signal toward the shadowed CPE/UE positions as visualized in Fig. \ref{fig:N78RISAssistedSetup}.

Here, a key challenge was that configuring the RIS directly from the KPIs experienced by the CPE was not feasible due to not being able to distinguish the RIS-oriented channel reportings since the network infrastructure has no RIS awareness. To address this, the RIS phase configuration was derived using an Adalm Pluto SDR and directional antenna operating as a receiver at the target UE positions to measure the RIS-assisted incident signal power from the AAU. The iterative optimization algorithm executed on the SDR produced the phase profile subsequently loaded onto the RIS controller. The validity of the derived configuration for the 5G link was confirmed by evaluating its effect on the received 5G signal via a spectrum analyzer interface, enabling a cross-validation between the SDR-computed configuration and the commercial network response. For the before/after comparison, a total of $8$ fixed UE positions within the NLoS region were selected. At each position, KPIs were measured sequentially in two states: RIS-off and RIS-on. This RIS-off/RIS-on toggle methodology isolates the RIS contribution from ambient channel variations. The performance gains introduced by the RIS are quantified in terms of the power-based KPIs: RSRP and SINR.

\section{Experimental Results} 
\label{sec:results}
This section presents the experimental results obtained from the three-phase measurement campaign, which are organized to reflect the progression from controlled indoor and outdoor experiments to real-world commercial network evaluations. 

\subsection{Indoor Results}
\begin{figure}[t]
  \centering
  \includegraphics[width=0.44\textwidth]{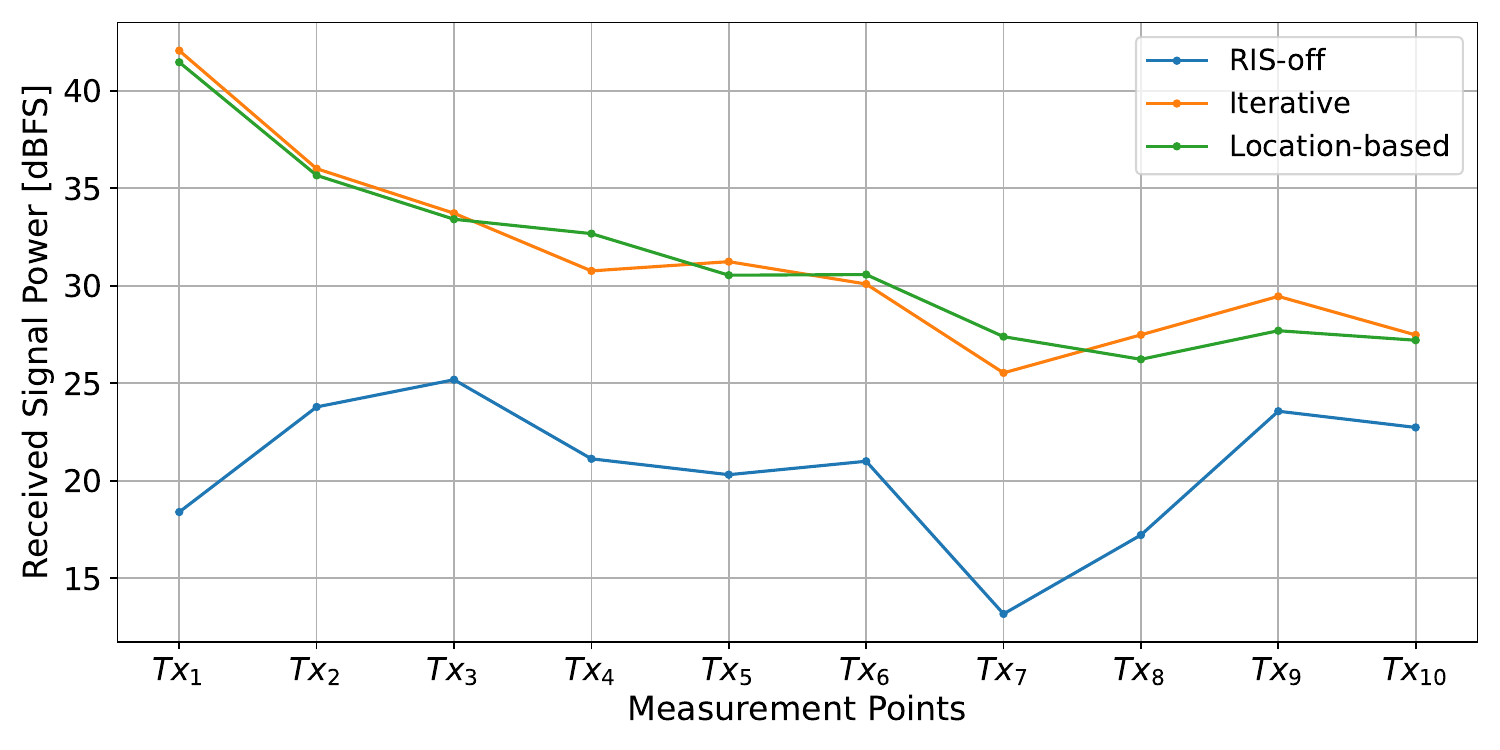}
  \caption{Indoor RIS-assisted communication measurement results}\vspace{-10pt}
  \label{fig:indoor_result}
\end{figure}

Following the description of the indoor measurement setup in the previous section, Fig. \ref{fig:indoor_result} presents the received signal power recorded across 10 distinct measurement locations for the transmitter ($Tx_1$ through $Tx_{10}$). Both iterative and location-based algorithms are employed to optimize the RIS configuration. As the figure illustrates, both approaches achieve comparable performance, delivering a substantial power gain over the baseline RIS-off state across all points. Additionally, the overall trend demonstrates expected path loss effects; as the distance between the transmitter and the RIS increases across the successive measurement points, the received signal power experiences a corresponding decline.

\begin{figure}[t]
  \centering
  \includegraphics[width=0.44\textwidth]{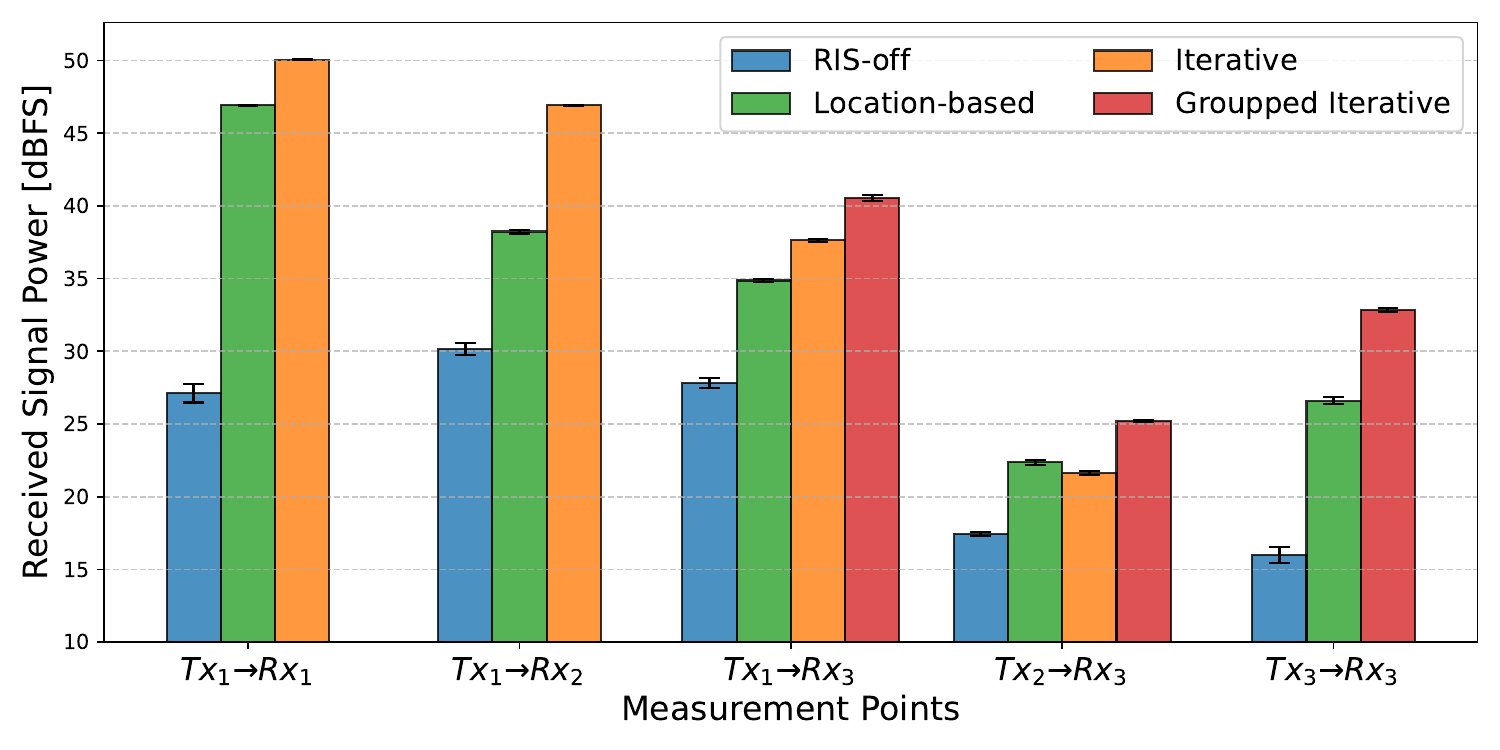}
  \caption{Outdoor RIS-assisted communication measurement results}\vspace{-10pt}
  \label{fig:outdoor_result}
\end{figure}

\subsection{Outdoor Results}
The measurement results for the outdoor RIS-assisted communication scenarios are presented in Fig. \ref{fig:outdoor_result}. Across all evaluated transmitter-receiver links, configuring the RIS yields a substantial improvement in received signal power compared to the baseline RIS-off state. For shorter link distances, specifically from $Tx_1$ to $Rx_1$ and $Rx_2$, the iterative algorithm achieves the highest received power, outperforming the location-based method. However, as the overall distance between the transceivers and the RIS increases (e.g., links involving $Tx_2$, $Tx_3$, or $Rx_3$), the effectiveness of the iterative method diminishes, eventually falling behind the location-based approach in the locations from $Tx_2$ to $Rx_3$. In these challenging long-distance scenarios, the adopted grouped iterative approach can be highly effective. By clustering RIS elements to enhance observable power variations during optimization, the grouped iterative method achieves the maximum signal power gain, demonstrating its robustness for far-field outdoor RIS applications.

\subsection{Commercial 5G Network Results}
\begin{figure}[t]
    \centering
    \subfigure[RSRP variation (top view)]{\includegraphics[width=0.22\textwidth]{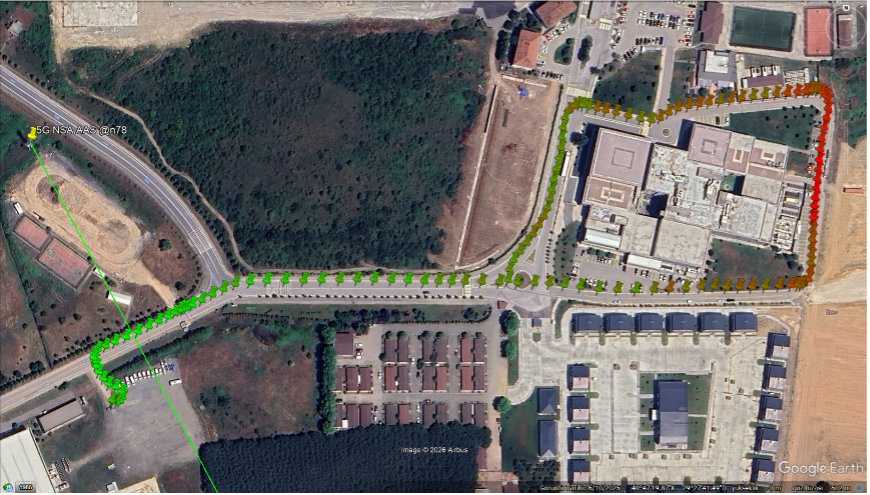}} \hfill
    \subfigure[SINR variation (top view)]{\includegraphics[width=0.22\textwidth]{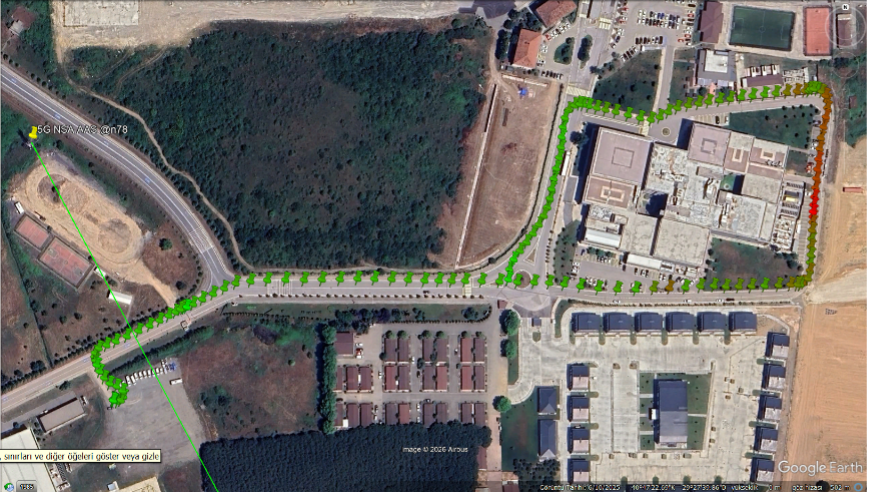}}
    \subfigure[RSRP values for time instances]{\includegraphics[width=0.22\textwidth]{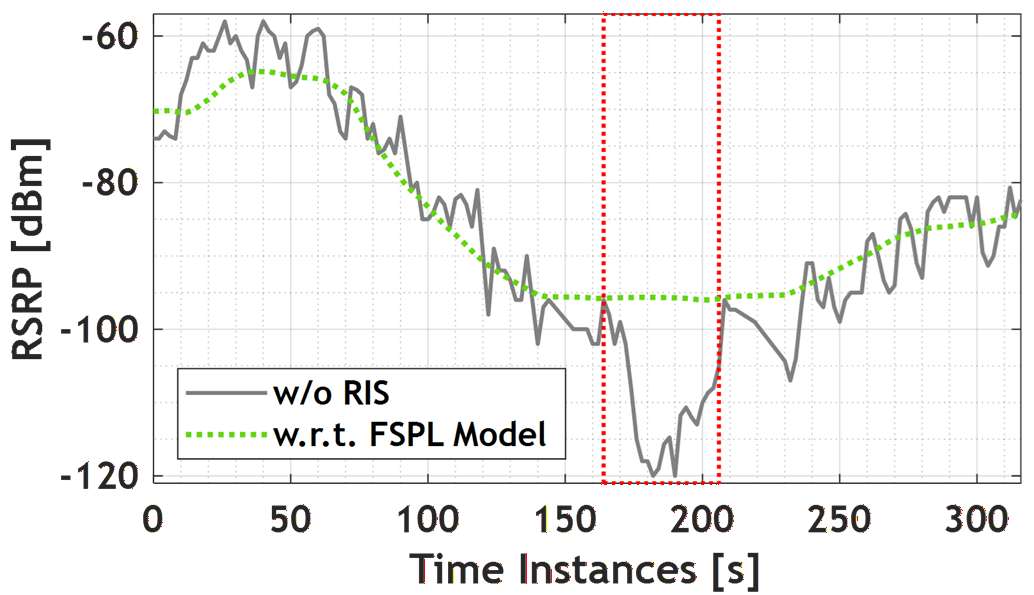}} \hfill
    \subfigure[SINR values for time instances]{\includegraphics[width=0.22\textwidth]{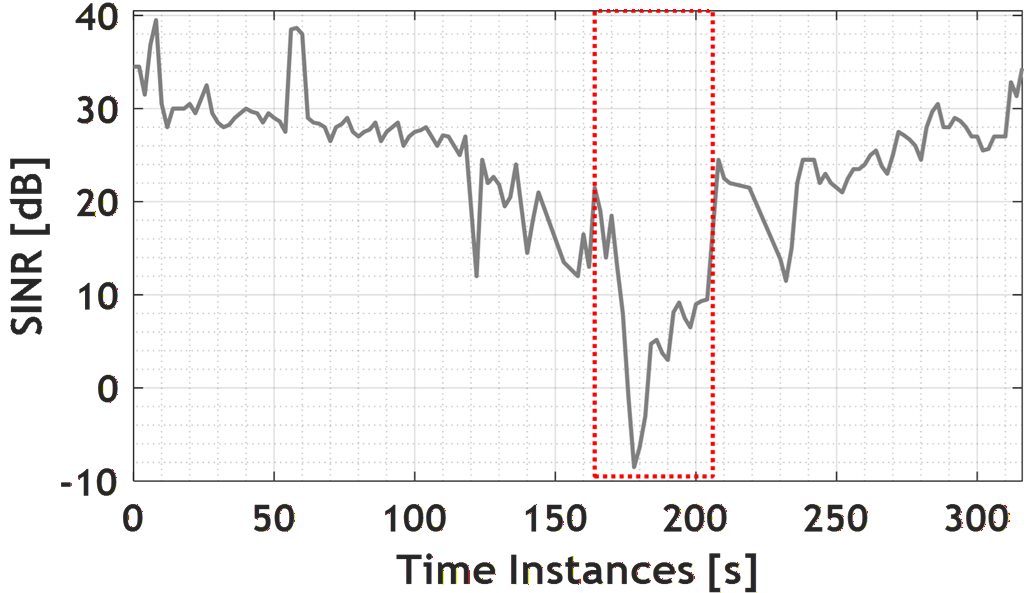}}
    \caption{Measured RSRP and SINR during the baseline (RIS-free) drive test campaign (higher values in green, lower in red).} \vspace{-10pt}
  \label{fig:kpi_baseline}
\end{figure}

Using the measurement methodology and infrastructure components described in Section III-C, signal strength-oriented performances were measured for both RIS inactive and active scenarios. As a result of the drive tests conducted to reveal the 5G wireless communication performance in the absence of RIS, as well as the coverage gaps, a total of $147$ time samples were recorded, including CPE's GPS coordinates, RSRP, RSRQ, RSSI, and SINR values. Due to space constraints, Fig.~\ref{fig:kpi_baseline} presents the spatial distributions of RSRP and SINR as the most representative indicators of 5G coverage.
Examining the top and bottom rows of Fig. \ref{fig:kpi_baseline}, it can be observed that, in addition to the path loss changes, both KPIs exhibit a dramatic drop due to building-induced blockage. The time intervals highlighted within the red dashed boxes in Fig. \ref{fig:kpi_baseline}-(c) and Fig. \ref{fig:kpi_baseline}-(d) correspond to CPE locations where this drop is most pronounced. To demonstrate that the observed dramatic effect is not caused by natural path loss, FSPL values are overlaid as a green dashed line alongside the RSRP results visualized in Fig. \ref{fig:kpi_baseline}-(c) for comparison purposes. While plotting this curve, an amplitude adjustment was applied to facilitate visual comparison.
These observations confirm the presence of a coverage hole attributable to physical blockage between the AAU and some UE locations. Based on this baseline assessment, the RIS was deployed on a dedicated outdoor stand within the measurement area, positioned to maintain LoS to the serving gNB and to redirect the incident signal toward the identified NLoS region. This deployment geometry ensures that the RIS establishes an alternative propagation path for UE positions that are otherwise shadowed.

\begin{figure}[t]
    \centering
    \subfigure[RSRP comparison]{\includegraphics[width=0.24\textwidth]{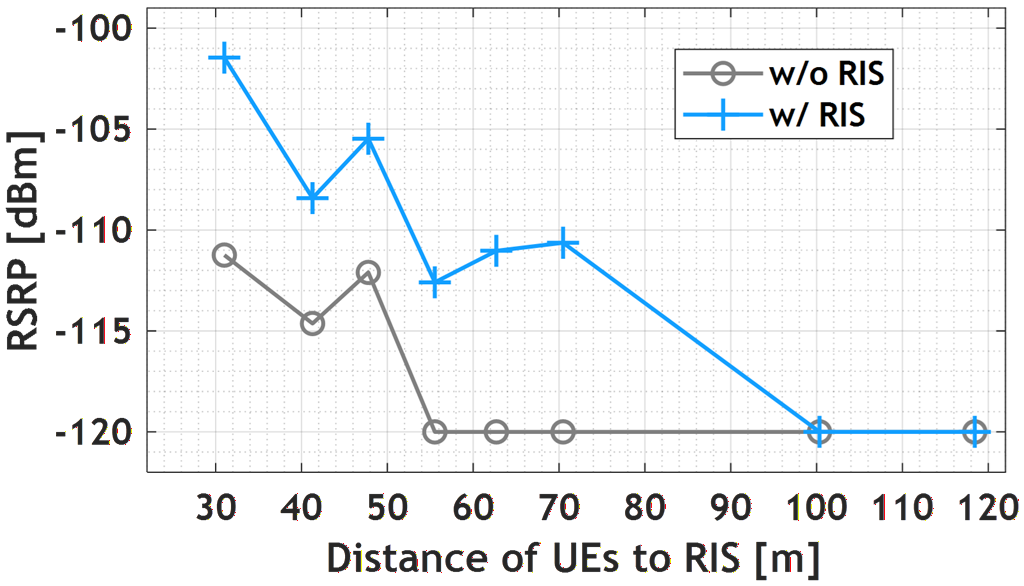}} \hfill
    \subfigure[SINR comparison]{\includegraphics[width=0.24\textwidth]{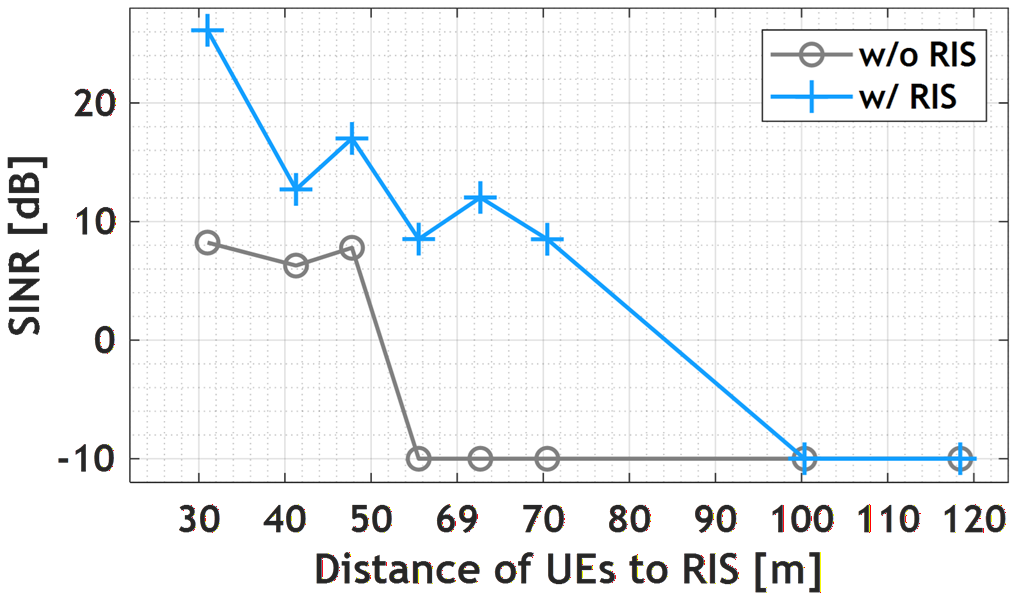}}
    \caption{Comparison of RSRP and SINR measured at fixed CPE/UE points}\vspace{-10pt}
  \label{fig:kpi_comparison}
\end{figure}

To demonstrate the performance gains achieved by the deployment of the N78 RIS prototype, the RIS phase configurations are optimized for $8$ different fixed UE locations by using the approach described in Section III-C. Using these configurations, both RIS-off and RIS-on performance at these points were measured in terms of power-based KPIs and presented in Fig. \ref{fig:kpi_comparison}. In the region identified as the coverage gap, the RSRP and SINR values observed in the absence of RIS are lower than those recorded during the drive test, moreover, starting from the 4th UE location, 5G service is interrupted and initial access to the wireless network cannot be established. The discrepancy between the KPIs observed during drive tests and fixed-point measurements can be attributed to the fact that, under mobility, time and spatial diversity together with beam tracking mechanisms enable the UE/CPE to sustain an already established connection even under degraded radio conditions, whereas at a fixed location the initial access procedure (e.g., SSB detection and PRACH-based random access) requires more stringent and stable signal quality thresholds, this behavior is consistent with the well-established practical observation that maintaining a connection is inherently more robust than establishing one, as also discussed in \cite{3GPPTS38300} and 
\cite{InitialAccess}.

In Fig. \ref{fig:kpi_comparison}, the RSRP and SINR values corresponding to scenarios where 5G wireless network access could not be established were set to $-120$ dBm and $-10$ dB, respectively, for visualization purposes. When comparing the KPI measurements obtained with the RIS active to those without RIS, RIS-induced RSRP and SINR gains are observed at the first three UE locations where 5G communication KPIs are calculated. While these gains vary across individual UE locations, their averages are approximately $8$ dB and $9$ dB, respectively. At the $4$th, $5$th, and $6$th UE locations, UEs that could not receive service without RIS are elevated to KPI levels sufficient to access 5G service with RIS support. At the last two examined points, the signal conditions in both cases remain insufficient to meet the requirements for initial access to the 5G network.

\section{Conclusions and Future Works}
This study presented a comprehensive experimental evaluation of a modular RIS prototype operating in the 5G N78 band. Through a three-phase methodology, including controlled indoor tests, outdoor long-range experiments, and deployment in a commercial 5G network, the practical performance of the RIS was assessed. Results demonstrated significant enhancements in signal quality, with average RSRP and SINR gains of approximately 8 dB and 9 dB, respectively, and enabled 5G service at previously inaccessible user locations. These findings confirm that RIS technology can effectively mitigate coverage gaps caused by physical blockages, even in mid-band deployments where scattering is prominent.

Future work will focus on optimizing RIS configurations in real-time using network-aware algorithms, integrating RIS with dynamic beam management in operational 5G networks, and conducting comprehensive downlink (DL) and uplink (UL) throughput measurements to complement the signal-level KPI analysis. In addition, from a theoretical perspective, intelligent algorithms will be developed to more efficiently identify coverage gaps in environments influenced by multiple 5G BSs, enabling more effective RIS deployment and configuration in complex network scenarios.

\section*{Acknowledgment}
This study has been supported by the 1515 Frontier Research and Development Laboratories Support Program of TÜBİTAK under Project 5229901 - 6GEN. Lab: 6G and Artificial Intelligence Laboratory. 
The authors would like to express their gratitude to Murat Yılmaz, and Ersan Algül from Turkcell, and Furkan Usta from TÜBİTAK for their supports regarding the network setup and experiments.

\bibliographystyle{IEEEtran}
\bibliography{references.bib}

@INPROCEEDINGS{yerliRIS,
 author={Kayrakl{\i}k, Sefa and Baș, Recep and {\c{C}}al{\i}{\c{s}}kan, Hasan O{\u{g}}uzhan and {\c{S}}ahino{\u{g}}lu, Samed and Erdo{\u{g}}an, Sercan and {\"U}nal, {\.I}lhami and H{\"o}kelek, {\.I}brahim and Nurdan, K{\i}van{\c{c}} and G{\"o}r{\c{c}}in, Ali},
  booktitle={2025 55th European Microwave Conference (EuMC)}, 
  title={N78 Frequency Band Modular {RIS} Design and Implementation}, 
  year={2025},
  volume={},
  number={},
  pages={795-798}}

@article{bjornson2019mMIMO,
  author={Bjornson, Emil and Van der Perre, Liesbet and Buzzi, Stefano and Larsson, Erik G.},
  journal={IEEE Wir. Commun.}, 
  title={Massive {MIMO} in Sub-6 {GHz} and {mmWave}: Physical, Practical, and Use-Case Differences}, 
  year={2019},
  volume={26},
  number={2},
  pages={100-108},
  doi={10.1109/MWC.2018.1800140}
}

@techreport{etsiRIS2025,
  author={{ETSI}},
  title={Reconfigurable Intelligent Surfaces {(RIS)}; Study on propagation models and use cases},
  institution={ETSI},
  number={GR RIS 003},
  year={2025}
}

@article{wu2019irs,
   author={Wu, Qingqing and Zhang, Rui},
  journal={IEEE Transactions on Wireless Communications}, 
  title={Intelligent Reflecting Surface Enhanced Wireless Network via Joint Active and Passive Beamforming}, 
  year={2019},
  volume={18},
  number={11},
  pages={5394-5409},
  doi={10.1109/TWC.2019.2936025}
}

@INPROCEEDINGS{kilinc2021channel,
  author={Kilinc, Fatih and Yildirim, Ibrahim and Basar, Ertugrul},
  booktitle={2021 55th Asilomar Conf. Signals, Systems, and Computers}, 
  title={Physical Channel Modeling for {RIS}-Empowered Wireless Networks in Sub-6 {GHz} Bands}, 
  year={2021},
  volume={},
  number={},
  pages={704-708},
  doi={10.1109/IEEECONF53345.2021.9723295}
}

@article{sang2023sub6,
  author={Sang, Jian and Zhou, Mingyong and Lan, Jifeng and Gao, Boning and Tang, Wankai and Li, Xiao and Jin, Shi and Basar, Ertugrul and Li, Cen and Cheng, Qiang and Cui, Tie Jun},
  journal={IEEE Transactions on Wireless Communications}, 
  title={Multi-Scenario Broadband Channel Measurement and Modeling for Sub-6 {GHz RIS}-Assisted Wireless Communication Systems}, 
  year={2024},
  volume={23},
  number={6},
  pages={6312-6329},
  doi={10.1109/TWC.2023.3330977}
}

@techreport{111_3gpp38104,
  author      = {{3GPP}},
  title       = {{NR; Base Station (BS) Radio Transmission and Reception}},
  institution = {3rd Generation Partnership Project (3GPP)},
  number      = {TS 38.104},
  year        = {2025}
}

@article{222_Basar2019RIS,
  author  = {E. Basar and M. Di Renzo and J. de Rosny and M. Debbah and M.-S. Alouini and R. Zhang},
  title   = {Wireless Communications Through Reconfigurable Intelligent Surfaces},
  journal = {IEEE Access},
  volume  = {7},
  pages   = {116753--116773},
  year    = {2019},
  doi     = {10.1109/ACCESS.2019.2935192}
}

@article{333_Wu2021RISTutorial,
  author  = {Q. Wu and S. Zhang and B. Zheng and C. You and R. Zhang},
  title   = {Intelligent Reflecting Surface-Aided Wireless Communications: A Tutorial},
  journal = {IEEE Transactions on Communications},
  volume  = {69},
  number  = {5},
  pages   = {3313--3351},
  month   = may,
  year    = {2021},
  doi     = {10.1109/TCOMM.2021.3051897}
}

@article{444_Dai2020RISPrototype,
  author  = {L. Dai and B. Wang and M. Wang and X. Yang and J. Tan and S. Bi and S. Xu and F. Yang and Z. Chen and M. Di Renzo and C.-B. Chae and L. Hanzo},
  title   = {Reconfigurable Intelligent Surface-Based Wireless Communications: Antenna Design, Prototyping, and Experimental Results},
  journal = {IEEE Access},
  volume  = {8},
  pages   = {45913--45923},
  year    = {2020},
  doi     = {10.1109/ACCESS.2020.2977772}
}

@article{555_Araghi2022Sub6RIS,
  author  = {A. Araghi and M. Khalily and M. Safaei and A. Bagheri and V. Singh and F. Wang and R. Tafazolli},
  title   = {Reconfigurable Intelligent Surface {(RIS)} in the Sub-6 {GHz} Band: Design, Implementation, and Real-World Demonstration},
  journal = {IEEE Access},
  year    = {2022},
  doi     = {10.1109/ACCESS.2022.3140278}
}

@ARTICLE{EA_Practical_2,
  author={Arslan, Emre and Dogukan, Ali Tugberk and Kilinc, Fatih and Coskun, Ahmet Faruk and Basar, Ertugrul},
  journal={IEEE Wireless Communications}, 
  title={Reconfigurable Intelligent Surface Identification in Mobile Networks: Opportunities and Challenges}, 
  year={2025},
  volume={32},
  number={4},
  pages={132-139},
  doi={10.1109/MWC.002.2400228}}

@ARTICLE{InitialAccess,
  author={Giordani, Marco and Polese, Michele and Roy, Arnab and Castor, Douglas and Zorzi, Michele},
  journal={IEEE Communications Surveys \& Tutorials}, 
  title={A Tutorial on Beam Management for {3GPP NR} at {mmWave} Frequencies}, 
  year={2019},
  volume={21},
  number={1},
  pages={173-196},
  doi={10.1109/COMST.2018.2869411}}

@techreport{3GPPTS38300,
  author       = {{3rd Generation Partnership Project (3GPP)}},
  title        = {{NR; Overall description; Stage-2}},
  institution  = {3GPP},
  type         = {Technical Specification},
  number       = {TS 38.300},
  year         = {2023},
  url          = {}}

@ARTICLE{iterative,
  author={Kayraklik, Sefa and Yildirim, Ibrahim and Hokelek, Ibrahim and Gevez, Yarkin and Basar, Ertugrul and Gorcin, Ali},
  journal={IEEE Open J. Commun. Soc.}, 
  title={Indoor Measurements for {RIS}-Aided Communication: Practical Phase Shift Optimization, Coverage Enhancement, and Physical Layer Security}, 
  year={2024},
  volume={5},
  number={},
  pages={1243-1255},
  doi={10.1109/OJCOMS.2024.3363423}}

@INPROCEEDINGS{BFTypeRISConfig,
  author={Ellingson, S. W.},
  booktitle={2021 IEEE 32nd Annual International Symp. Personal, Indoor and Mobile Radio Commun. (PIMRC)}, 
  title={Path Loss in Reconfigurable Intelligent Surface-Enabled Channels}, 
  year={2021},
  volume={},
  number={},
  pages={829-835},
  doi={10.1109/PIMRC50174.2021.9569465}}

\end{document}